\documentclass{emulateapj}
\usepackage{epsfig,lscape}
\def\lap{\lower.5ex\hbox{$\; \buildrel < \over \sim \;$}}
\def\gap{\lower.5ex\hbox{$\; \buildrel > \over \sim \;$}}

\def\ergcm2s{${\rm erg\ cm^{-2}\ s^{-1}}$}

\def\ergscm2s{${\rm erg\ cm^{-2}\  s^{-1}}$}

\def\cm-2{${\rm cm^{-2}}$}

\begin{document}

\title{The ACS Nearby Galaxy Survey Treasury IV. The Star Formation History of
NGC 2976}

\author{Benjamin F. Williams\altaffilmark{1},
Julianne J. Dalcanton\altaffilmark{1},
Adrienne Stilp\altaffilmark{1},
Karoline M. Gilbert\altaffilmark{1},
Rok {Ro{\v s}kar}\altaffilmark{1},
Anil C. Seth\altaffilmark{2},
Daniel Weisz\altaffilmark{3},
Andrew Dolphin\altaffilmark{4},
Stephanie M. Gogarten\altaffilmark{1},
Evan Skillman\altaffilmark{3},
Jon Holtzman\altaffilmark{5}
}

\altaffiltext{1}{Department of Astronomy, Box 351580, University of Washington, Seattle, WA 98195; ben@astro.washington.edu; jd@astro.washington.edu; stephanie@astro.washington.edu; roskar@astro.washington.edu}
\altaffiltext{2}{CfA Fellow, Harvard-Smithsonian Center for Astrophysics, 60 Garden Street, Cambridge, MA 02138; aseth@cfa.harvard.edu}
\altaffiltext{3}{Department of Astronomy, University of Minnesota, 116 Church
St. SE, Minneapolis, MN 55455; dweisz@astro.umn.edu; skillman@astro.umn.edu}
\altaffiltext{4}{Raytheon, 1151 E. Hermans Road, Tucson, AZ 85706; dolphin@raytheon.com}
\altaffiltext{5}{Department of Astronomy, New Mexico State University, Box
30001, 1320 Frenger St., Las Cruces, NM 88003; holtz@nmsu.edu}


\begin{abstract}

We present resolved stellar photometry of NGC~2976 obtained with the
Advanced Camera for Surveys (ACS) as part of the ACS Nearby Galaxy
Survey Treasury (ANGST) program.  The data cover the radial extent of
the major axis of the disk out to 6~kpc, or $\sim$6 scale lengths.
The outer disk was imaged to a depth of $M_{F606W}\sim 1$, and an
inner field was imaged to the crowding limit at a depth of
$M_{F606W}\sim -1$.  Through detailed analysis and modeling of these
CMDs we have reconstructed the star formation history of the stellar
populations currently residing in these portions of the galaxy,
finding similar ancient populations at all radii but significantly
different young populations at increasing radii. In particular,
outside of the well-measured break in the disk surface brightness
profile, the age of the youngest population increases with distance
from the galaxy center, suggesting that star formation is shutting
down from the outside-in.  We use our measured star formation history,
along with H~I surface density measurements, to reconstruct the
surface density profile of the disk during previous epochs.
Comparisons between the recovered star formation rates and
reconstructed gas densities at previous epochs are consistent with
star formation following the Schmidt law during the past 0.5 Gyrs, but
with a drop in star formation efficiency at low gas densities, as seen
in local galaxies at the present day.  The current rate and gas
density suggest that rapid star formation in NGC~2976 is currently in
the process of ceasing from the outside-in due to gas depletion.  This
process of outer disk gas depletion and inner disk star formation was
likely triggered by an interaction with the core of the M81 group
$\gap$1~Gyr ago that stripped the gas from the galaxy halo and/or
triggered gas inflow from the outer disk toward the galaxy center.

\end{abstract}

\section{Introduction}\label{intro}

Bursts of star formation play a significant and important role in the
evolution of galaxies, particularly at low masses, such as dwarf
galaxies.  Star formation histories (SFHs) of dwarfs in the Local
Group show strong evidence for past or current bursts
\citep{mateo1998,dohm-palmer2002,dolphin2005,young2007,cole2007}. The
SFHs of the non-tidal dwarfs in the M81 group are equally diverse
\citep{weisz2008}.  Recent observations of the entire local volume
within 11 Mpc also indicate that bursts are an essential mode of star
formation in low mass galaxies \citep{lee2009}.  Furthermore,
numerical simulations suggest that episodic bursts of star formation
should be common in low-mass galaxies, as gas is expelled by
supernovae and then falls back into the galaxy
\citep[e.g.][]{stinson2007}.

While there is little question that bursts are important in the
formation and evolution of low-mass galaxies, most of the stellar mass
in dwarfs may not form in bursts.  For example, several studies
suggest that dwarf irregular galaxies have relatively constant star
formation histories \citep{gallagher1984,greggio1993}.  Even though
bursts clearly occur, most recent measurements suggest they are
responsible for the production of only about a quarter of the stellar
mass in dwarfs \citep{lee2009}. Recent studies of nearby low mass
starbursts have also put constraints on their durations, suggesting
they typically last 200--400 Myr \citep{mcquinn2009}, in agreement
with results from numerical simulations.

Much of the recent progress made in our understanding of the role of
star formation bursts in the evolution of low-mass galaxies is due to
the availability of deep resolved stellar photometry from the {\it
Hubble Space Telescope (HST)}.  Using such deep photometry, it
possible to study galaxies using techniques previously available only
within the Local Group.  Rather than inferring details of evolution
from integrated properties such as scale length, color, or surface
brightness, the star formation history (SFH) of the stars in the
galaxy can be determined by fitting the distribution of stars in
color-magnitude diagrams (CMDs) with model distributions determined
from stellar evolution isochrones, \citep[see e.g.][for Local Group
examples]{holtzman1999,dolphin2000b,williams2002,dohm-palmer2002,harris2004,dolphin2005}.

Detailed measurements of SFHs can be compared to the results of
numerical simulations, which now probe the properties of the stellar
populations they produce, to shed light on the physical processes
responsible for the morphological structures we observe.  Such
simulations are beginning to show the importance of galaxy
interactions in the evolution of disks \citep{governato2007}, of
internal dynamical interactions in distributing stars of different
ages throughout the disk \citep{roskar2008}, and of galaxy mass,
feedback and gas availability in the process of star formation
\citep{stinson2007}.  However, the simulations that show the most
detailed features do not yet include environmental effects and cover
limited ranges of galaxy mass.

NGC~2976, which has $M_B = -17$, $W_{20} = 165$ km s$^{-1}$, $(m-M)_0
= 27.76\pm0.23$, $A_V=0.23$, and a morphology of SAc pec,
\citep{simon2003,karachentsev2002,schlegel1998}, is an excellent
specimen for studying the effects of bursts of star formation and
environment on the evolution of low-mass galaxies.  This peculiar
low-mass disk galaxy lies on the outskirts of the M81 group, outside
of the \citet{yun1994} H~I maps.  The inner disk of the galaxy
contains many young stars, has a sharp truncation edge
\citep{simon2003}, appears to have a barred potential
\citep{spekkens2007}, and may be inside of a larger spheroidal halo
\citep{bronkalla1992}.  The galaxy's gas content and ongoing star
formation \citep{bigiel2008,leroy2008} are affecting its morphology.
In order to investigate this impact, as well as potential effects of
the M81 group environment, we have performed a systematic study of the
stellar populations as a function of galactocentric distance in this
peculiar galaxy.  Our results suggest NGC~2976 is undergoing a
transition from a burst of star formation to a more quiescent state,
and thus we have a unique opportunity to study the final stages of a
common mode of star formation.

Herein we report the SFH, measured via CMD fitting, of several regions
of NGC~2976.  We look for trends with radius, analyzing regions of
constant crowding and differential extinction.  We show that the
ancient population is consistent with being similar over all radii.
However, the age of the young population appears to increase sharply
with radius beyond the disk break.  By reconstructing the past gas
surface density profile, we show that such an abrupt change in the age
distribution is likely to be due to the depletion of gas in the
NGC~2976 outer disk, possibly caused by an interaction with the core
of the M81 group.

The paper is organized as follows.  Section 2 details our data
acquisition and reduction techniques. Section 3 compares and contrasts
the recovered SFHs of the different regions. Section 4 discusses and
interprets potential explanations for the differences in a galaxy
evolution context. Finally, section 5 summarizes our main conclusions.
We assume $(m-M)_0 = 27.76\pm0.23$ \citep{karachentsev2002} for
conversions of angular measurements to physical distances and an
inclination angle $i$=64.5$^{\circ}$ \citep{deblok2008} for surface
density measurements.  We adopt a WMAP \citep{dunkley2009} cosmology
for all conversions between time and redshift.

\section{Data Acquisition, Reduction, and Analysis}\label{data}

\subsection{Acquisition}\label{acquisition}

\subsubsection{Outer Disk Field}\label{outerdata}

\begin{figure}
\centerline{\psfig{file=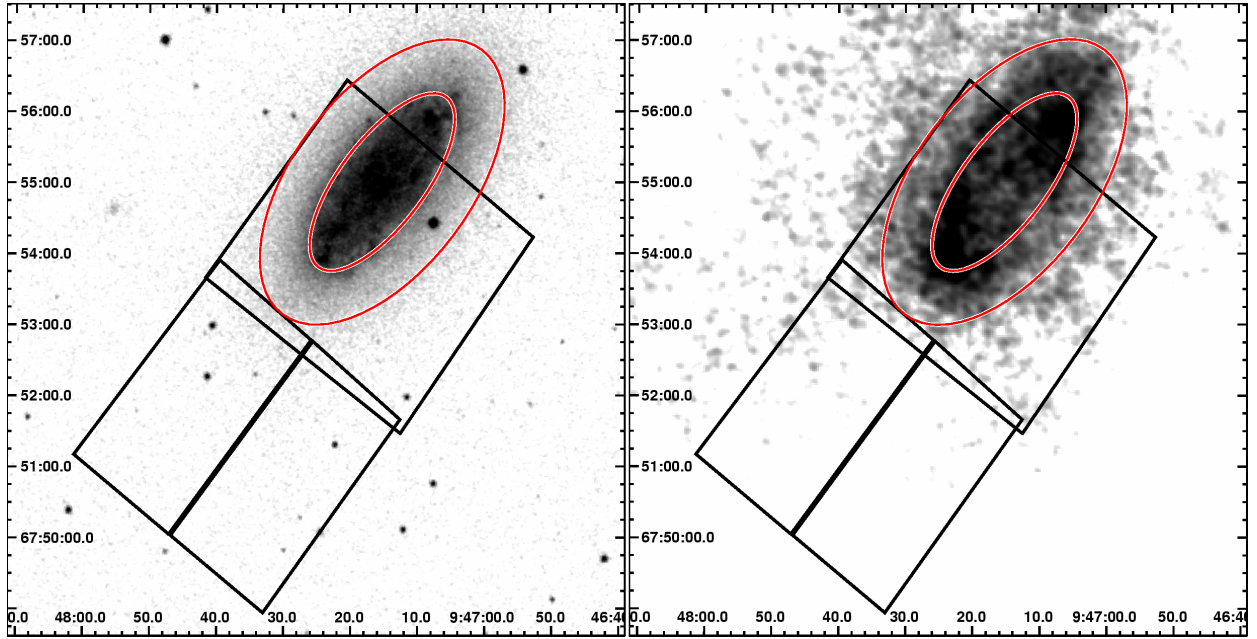,width=6.5in,angle=0}}
\caption{The locations of our NGC~2976 fields are shown on a DSS image
(left) and the H~I Nearby Galaxy Survey (THINGS) image (right). The
inner field shows no chip gap because we dithered over the chip gap in
this field.  Ellipses denote our region boundaries (see
\S~\ref{data}).}
\label{field_loc}
\end{figure}

From 2006-Dec-27 to 2007-Jan-10, we observed a field covering the
inner part of the NGC~2976 disk located at R.A.~(2000) =146.902554
(09:47:36.61), decl.~(2000) = +67.85705 (+67:51:25.4) with a rotation
angle PA\_V3=51.35.  We also observed a field in the outskirts of the
NGC~2976 disk located at R.A.~(2000) =146.820171 (09:47:17.21),
decl.~(2000) = +67.898886 (+67:53:55.99) with a rotation angle
PA\_V3=52.75, taking advantage of the lower stellar density to image
this outer field more deeply. Figure~\ref{field_loc} shows outlines of
the field locations, which cover the nucleus of the galaxy and extend
to 5.7$'$ (6 kpc at NGC~2976) along the major axis.  This distance
corresponds to $\sim$6 scale lengths (mean of inner and outer disk)
using the measured scale length from the isophotes on our F814W ACS
images of the inner disk.  If the disk/halo components of NGC~2976
scale as they do with larger disk galaxies, such as M31 and M81, the
disk population should dominate out to at least 6 scale lengths,
suggesting that our deep field is mainly probing the outer disk
population.

In the outer field, we obtained 1 orbit of exposures with the ACS
\citep{ford1998} through the F475W (SDSS g' equivalent) filter, 7
full-orbit exposures through the F606W (wide $V$) filter, and 10
full-orbit exposures through the F814W ($I$ equivalent) filter.  These
data totaled 2418~s, 18716~s and 27091~s of exposure time in F475W,
F606W, and F814W, respectively.  In the inner field, we observed for 2
orbits, obtaining 2 dithered exposures through the F475W (wide $B$),
F606W (wide $V$), and F814W ($I$ equivalent) filters. Deeper
observations for this field would not have been efficient, as the
inner regions of the galaxy are too crowded to resolve faint
stars. However, we dithered to cover the ACS chip gap so that we did
not miss any of the main body of the galaxy.  These data totaled
1570~s and 1596~s and 1622~s of exposure time in F475W, F606W, and
F814W, respectively. All routine image calibration, including bias
corrections and flat-fielding, were performed by the {\it HST}
pipeline, OPUS version 2006\_6, CALACS version 4.6.1.

\subsection{Reduction}\label{reduction}

Photometry and artificial star tests were made as part of the ANGST
survey pipeline.  What follows is a brief description of the
technique; readers are referred to \citet{dalcanton2009} for a fuller
description.  

We initially processed the images by running the {\tt multidrizzle}
task within PyRAF \citep{koekemoer}, which was used only to flag the
cosmic rays in the individual images and to produce a reference
coordinate system for the photometry.  Once the cosmic rays were
flagged, the photometry was measured simultaneously for all of the
objects in the uncombined images using the software package
DOLPHOT~1.0 \citep{dolphin2000} including the ACS module.  This
package is optimized for measuring photometry of stars on ACS images
using the well-characterized and stable point spread function (PSF),
calculated with
TinyTim.\footnote{http://www.stsci.edu/software/tinytim/} The software
fits the PSF to all of the stars in each individual frame to find PSF
magnitudes.  It then determines and applies the aperture correction
for each image using the most isolated stars, corrects for the charge
transfer efficiency of the ACS detector, combines the results from the
individual exposures, and converts the measured count rates to the
VEGAmag system.

The DOLPHOT output was then filtered to only allow objects classified
as stars with signal-to-noise $>$6 in both filters.  The list was
further culled using sharpness ($|F606W_{sharp} + F814W_{sharp}| <
0.27$) and crowding ($F606W_{crowd} + F814W_{crowd} < 0.1$).  The
sharpness cut was chosen based on the distribution of values in the
original catalog.  The crowding parameter gives the difference between
the magnitude of a star measured before and after subtracting the
neighboring stars in the image.  When this value is large, it suggests
that the star's photometry was significantly affected by crowding, and
we therefore exclude it from our catalog.  We also considered quality
cuts based on the $\chi$ values, but rejected them when a correlation
was found between $\chi$ and the local background.  Our final star
catalogs for the fields contained 96787 and 248864 stars detected in
both the F606W and F814W filters for the outer and inner fields,
respectively.  We detected 14142 and 112756 stars in the F475W and
F814W filters in the outer and inner fields, respectively.

Finally, a series of 2 million artificial star tests were run on each
field.  Model PSF stars were added to the data covering the full range
of color, magnitude, and position space measured in the stellar
photometry and including test stars up to 1 magnitude fainter than the
faintest measured stars.  The photometry was then measured by DOLPHOT
again to determine the likelihood that a star was recovered, and if
so, the offsets between its true and measured color and magnitude.

\subsection{CMD Fitting}\label{fitting}

We measured the star formation rate and metallicity as functions of
age, using the MATCH package \citep{dolphin2002}. The entire observed
CMD is fitted by populating the stellar evolution models of
\citet[][with updated AGB models from
\citealp{marigo2008}]{girardi2002} with a given initial mass function
(IMF), finding the distance modulus, extinction, and linear
combination of ages and metallicities that best fit the observed color
and magnitude distribution \citep[see details in][]{dolphin2002}.  A
full discussion of the choice of fitting software and stellar
evolution models used for the ANGST project is given in
\citet{williams2008}.

\subsubsection{Field Division}\label{division}

\begin{figure}
\centerline{\psfig{file=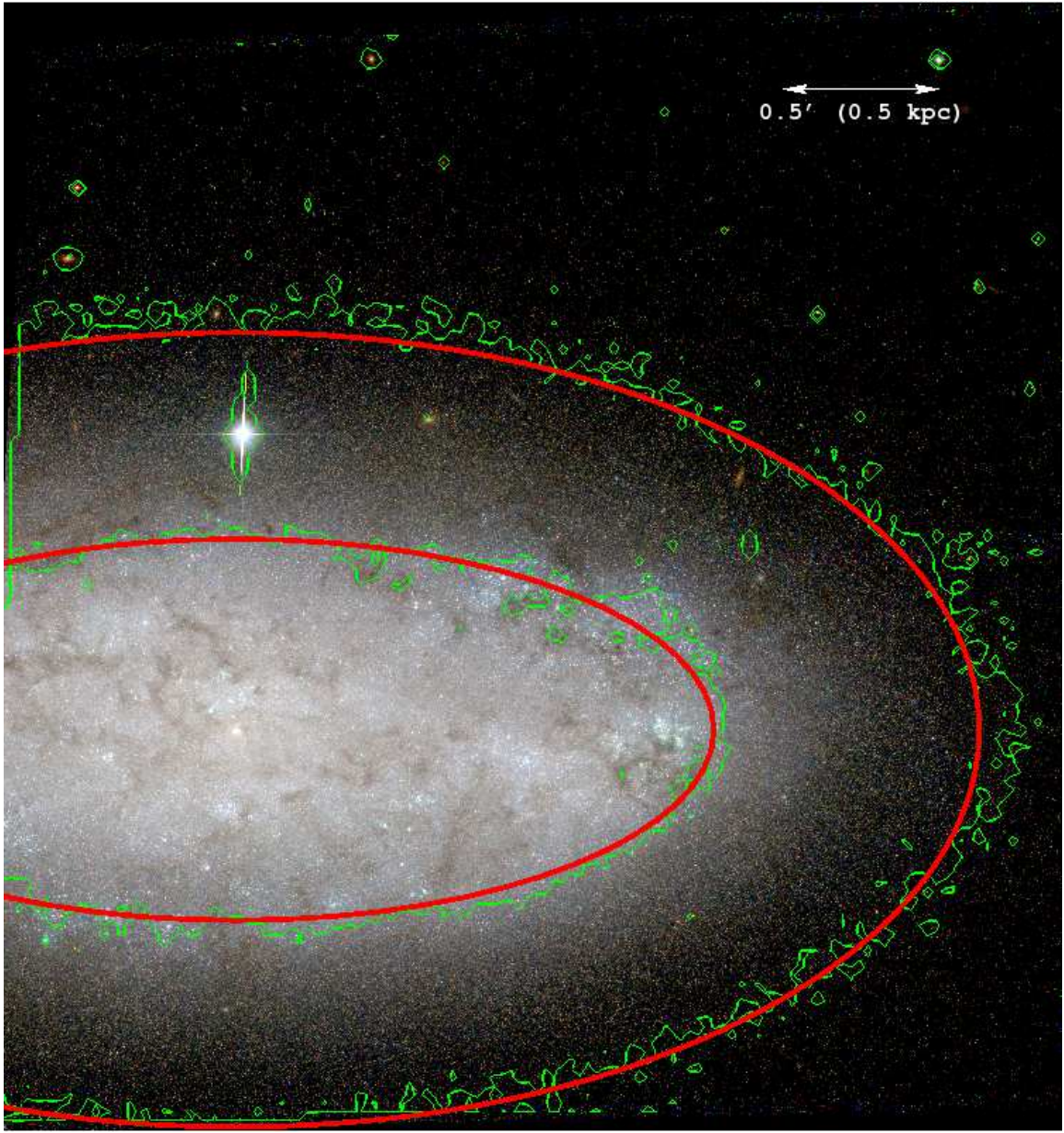,width=6.0in,angle=0}}
\caption{Our division of the inner field into 3 regions is shown (see
\S~\ref{division}).  Thin white lines show the contours of constant
surface brightness.  Thick white lines show the ellipses used to
cleanly sort stars into their appropriate regions.  The scale bar in
the upper right corner shows 0.5$'$, or 0.5~kpc at the distance of
NGC~2976.  The ellipses are centered at R.A. (J2000) = 09:47:15.185,
decl. (J2000) = +67:55:00.59, and they have semi-major axes of 90$''$
and 140$''$ and semi-minor axes of 36$''$ and 75$''$.}
\label{ellipses}
\end{figure}

\begin{figure}
\centerline{\psfig{file=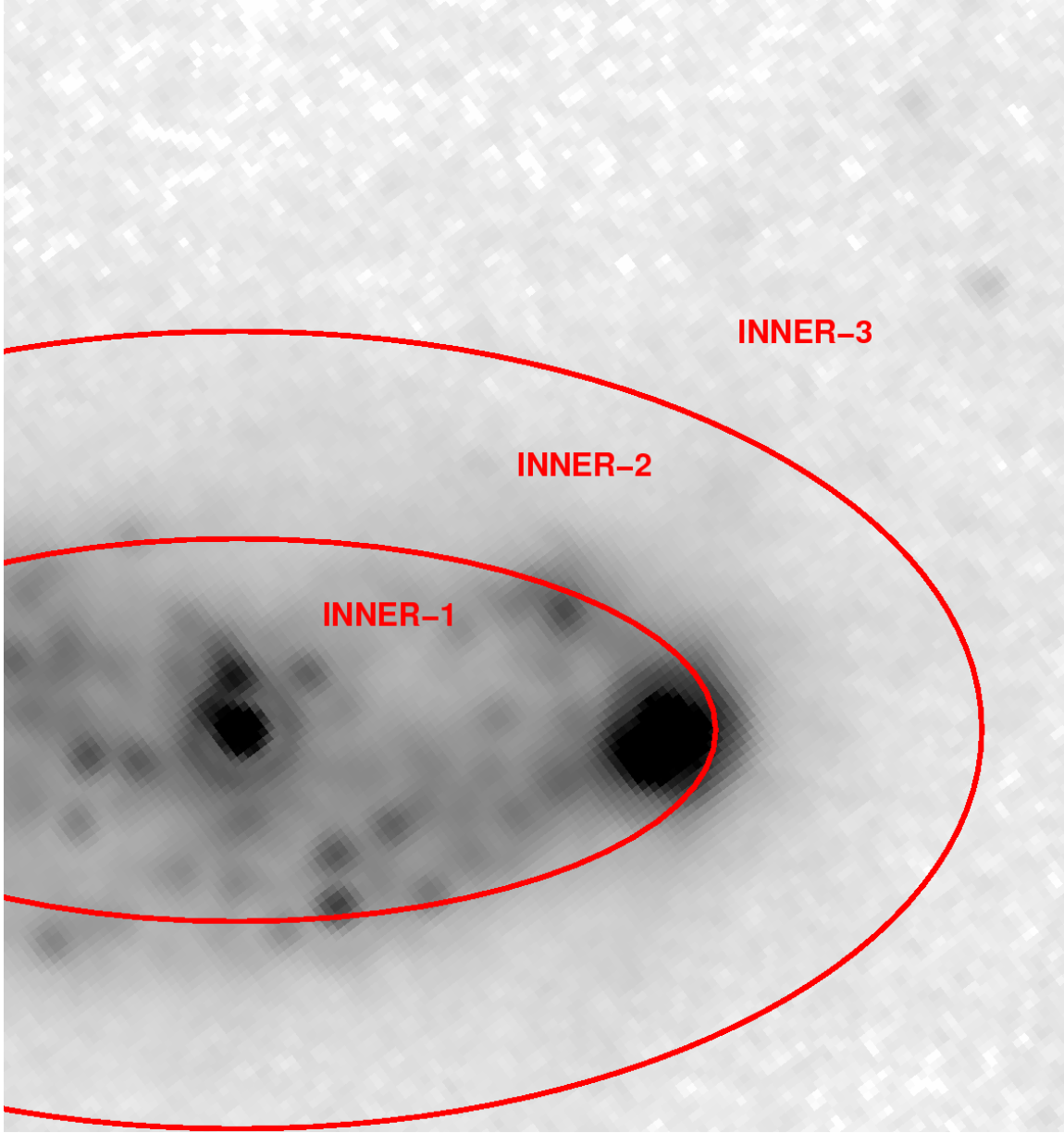 ,width=6.0in,angle=0}}
\caption{Our division of the galaxy marked with ellipses superimposed
on a 24$\mu$m {\it Spitzer} image (same dimensions as in
Figure~\ref{ellipses}).  The star formation is confined to the inner
ellipse (see \S~\ref{division}).}
\label{ir_ellipses}
\end{figure}

\begin{figure}
\centerline{\psfig{file=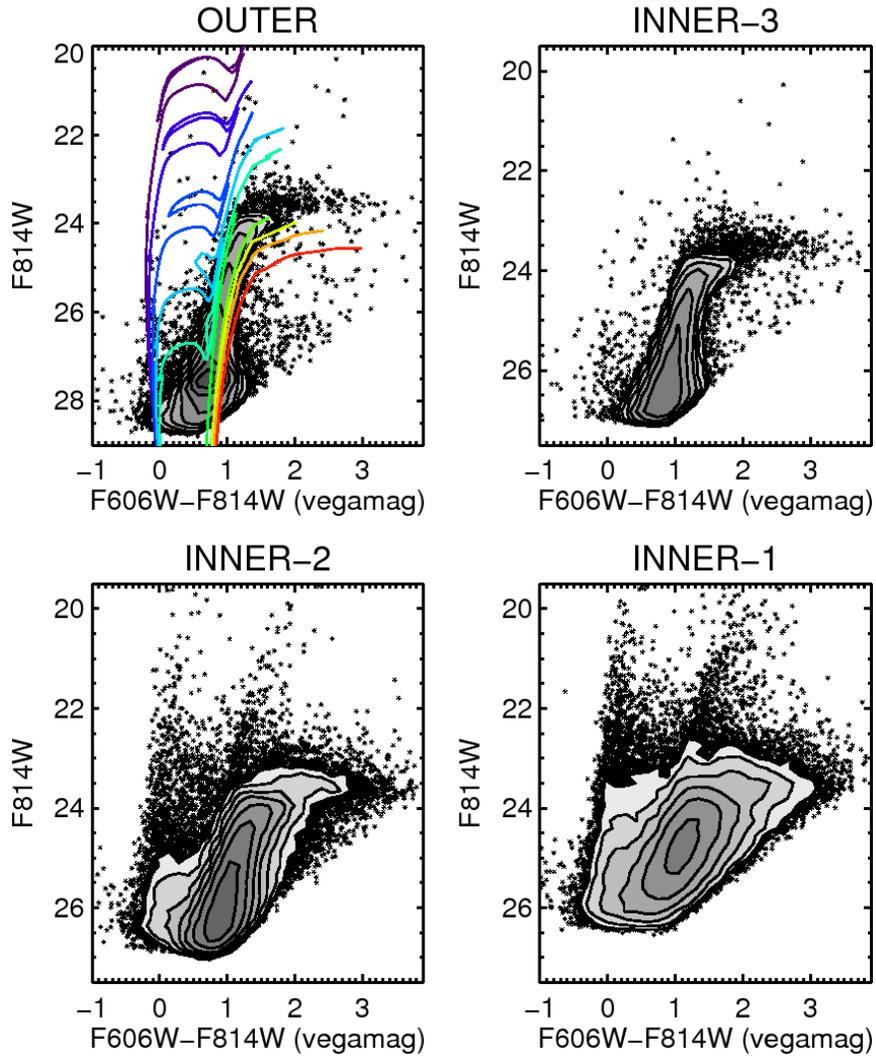,width=5.0in,angle=0}}
\caption{CMDs of the four regions (see \S~\ref{division}). Contours
show the density of points where they would otherwise saturate the
plot.  Levels are 1, 2, 4, 8, 16, 32, 64, and 128 thousand points
mag$^{-2}$. Lines in the upper-left panel show a small subset of
isochrones for reference \citep[][shifted assuming $A_V=0.1$ and
$(m-M)_0 = 27.75$]{marigo2008}. The 10 isochrones shown are (from blue
to red): [M/H]=-0.4 and log(age)~=~7.3,7.6,8.0,8.3,8.6, followed by
log(age)=10.0 and [M/H]~=~-1.3,-0.7,-0.4,-0.2,0.0, respectively.}
\label{cmds}
\end{figure}

As can be seen in Figure~\ref{ellipses}, NGC~2976 shows strong radial
variations in its surface brightness, dust content, and current star
formation rate.  We therefore choose to analyze the galaxy in a series
of annuli.  While the outer disk field does not show signs of
structure that would warrant the division of the field into
subregions, the inner disk field clearly has a strong gradient in
surface brightness.  This gradient causes the photometric completeness
and error statistics to vary significantly with position.  Following
the galaxy's isophotes, we divided the inner field into 3 regions,
shown in Figure~\ref{ellipses}.  The inner ellipse contains the
crowding-limited, high surface brightness inner galaxy and completely
covers the area interior to the known break in NGC~2976's surface
brightness profile at a radius of $\sim$1.2 kpc \citep[inner
$r_s\sim$1.3~kpc, outer $r_s\sim$0.6~kpc;][]{simon2003}.  This area
also contains most of the structured dust content according to the
24$\mu$ {\sl Spitzer} image shown in Figure~\ref{ir_ellipses}.  The
outer ellipse marks the transition from this inner region to the
uncrowded outer portions of the field.  These 2 ellipses delineate 3
regions in the inner field. Along with the outer field, the field
edges designate 4 separate regions of the galaxy for analysis, which
we labeled from the inside-out as INNER-1, INNER-2, INNER-3, and
OUTER.  INNER-1 contains the disk break and the dusty, crowded areas
within.  INNER-2 is outside the disk break and high dust content but
still contains a high density of stars with significant crowding, and
INNER-3 is outside of the crowding-limited area.  There is significant
overlap in the projected radii probed by the INNER-3 and OUTER
regions.  The corresponding CMDs are shown in Figure~\ref{cmds}.

With our artificial star tests, we determined the 50\% completeness
magnitudes of our 4 regions, given in Table~\ref{depth}.  We applied
these magnitude cuts to the data fitted by MATCH.

\subsection{Fitting Parameters}\label{binning}

To model the full CMD, we must adopt values for the binary fraction,
IMF slope, the area of the CMD to include in the fit, the approximate
distance and mean extinction to the stars in the field, and the
binning of the stellar evolution models in time and metallicity. Below
we discuss how we chose these parameters and how the choices impact
our results.

For fitting purposes, we assumed a binary fraction of 0.35 and a
\citet{salpeter1955} IMF when populating the model isochrones.  As has
been shown by other studies using this technique
\citep[e.g.,][]{williams2007,barker2007}, our choice of IMF does not
affect the relative star formation rates in the SFH, since our data do
not probe the main sequence at low masses, where the IMF differs from
a single power law \citep[e.g.][]{kroupa2002}. Once the fits are
complete, we can normalize the output star formation rates to other
IMFs whenever necessary.

We initially allowed the distance modulus to range from 27.6 to 28.0
and the extinction to range from $A_V=0.0$ to $A_V=0.4$ in the OUTER
region and from $A_V=0.0$ to $A_V=1.2$ in the INNER regions.  Within
these ranges, we allowed MATCH to determine the systematic errors that
result from small changes to these parameters and to optimize the
overall CMD fit, even in the presence of localized deficiencies in the
model isochrones.  However, once these systematic uncertainties were
characterized, we reran our fits with the distance for all fields
fixed to the best-fit value for the deep OUTER data ($m-M_0$=27.75).
The resulting SFHs were indistinguishable from those obtained when
distance was left as a free parameter.

The obvious dust visible in our inner disk field suggests that a
single foreground extinction value cannot properly account for the
effects of dust in our CMDs.  We have attempted to minimize the
effects of this problem in two ways.  First, we have divided the field
into separate regions with differing amounts of dust (e.g.,
Figure~\ref{ellipses}).  The radial variation in internal extinction
is evident in the foreground extinction values measured for each
region by MATCH, which drop from $A_V = 0.56\pm0.07$ within the
INNER-1 regions to $A_V=0.34\pm0.07$ and $A_V=0.21\pm0.07$ within the
INNER-2 and INNER-3 regions respectively.  The preferred extinction
falls even further in the OUTER region, down to $A_V = 0.10\pm0.04$,
which is below the value in the Schlegel maps of Galactic extinction.
This slight discrepancy corresponds to a color difference of
$\sim$0.03 mag.  Such a slight difference is less than the size of the
color bins in our Hess diagrams (0.05 mag).  This type of discrepancy
emphasizes the importance of fitting the overall distance and
extinction as free parameters in order to compensate for such small
discrepancies between the data and the model isochrones.

Second, we performed two fits to each of the INNER regions.  One fit
included 0.8 mag of differential extinction for ages $>$100 Myr, and
the other fit included no differential extinction for ages $>$ 100
Myr.  By default, MATCH includes an additional $\sim$0.5 mag of
differential extinction for ages $<$100 Myr. We found that including
differential extinction for ages $>$100 Myr decreased the fit quality
for the INNER-2, INNER-3 , and OUTER regions.  Therefore, we accepted
the fits that included no differential extinction for ages $>$ 100 Myr
for the these regions. For the INNER-1 region, we accepted only the
fits with 0.8 mag of differential extinction. Including this amplitude
of differential extinction improved the fit and brought the measured
foreground extinction value to $A_V = 0.46\pm0.06$.

\begin{figure}
\centerline{\psfig{file=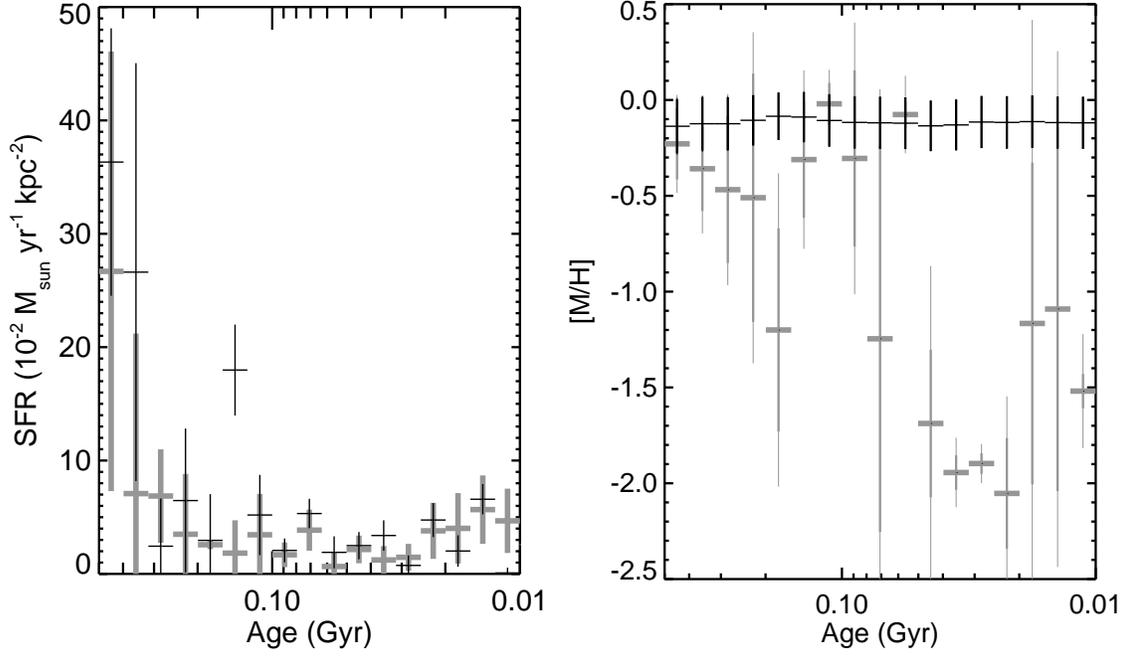,width=6.0in,angle=0}}
\caption{{\it Left:} The recent SFH for the INNER-1 region at full
time resolution assuming metallicity stays constant or increases with
time (thin black) and making no assumptions about enrichment (thick
gray).  The results agree within the uncertainties for all times
except the 10-12.5 Myr and 125-160 Myr bins. {\it Right:} The recent
metallicity history assuming metallicity stays constant or increases
with time (thin black) and making no assumptions about enrichment
(thick gray).  Note that the age distribution is not strongly affected
by the metallicity distribution at these ages (see \S~\ref{fitting}).}
\label{recent_testzinc}
\end{figure}

When fitting the CMD of the shallower and heavily reddened photometry
of the INNER-1 region, we further limited the number of free
parameters by imposing an ``increasing metallicity'' constraint on the
fit, whereby the metallicity of the population was not allowed to
decrease with time (within the measured errors). Without this
constraint, the best-fitting SFH was found to be dominated by stars
with solar metallicity at the oldest ages in order to fit the reddest
and most heavily extincted red giants.  Furthermore, without the
constraint, the best-fit metallicity fluctuated from [M/H]=-0.6 to
[M/H]=-0.1 then to [M/H]=-1.8 all in $<$1 Gyr.  Imposing the
metallicity constraint resulted in a lower fit quality, but it avoided
unphysical solutions, such as order of magnitude fluctuations in
metallicity on short ($\lap$100~Myr) timescales.  With the constraint
in place, the best-fitting metallicity for the young population was
found to be -0.12$\pm$0.14 and stayed constant for the past Gyr.  The
effects of the constraint on the recent SFH are shown in
Figure~\ref{recent_testzinc}.  The results with and without the
constraint agree within the uncertainties for all but two 0.1 dex time
bins. Thus, the metallicity constraint yielded a more physically
plausible metallicity (with no wild fluctuations) while having minimal
impact on the results of the age distribution of the young population.
This lack of sensitivity to the metallicity constraint is likely due
to the fact that the color of young stars on the upper main sequence
are not very sensitive to metallicity.

\begin{figure}
\centerline{\psfig{file=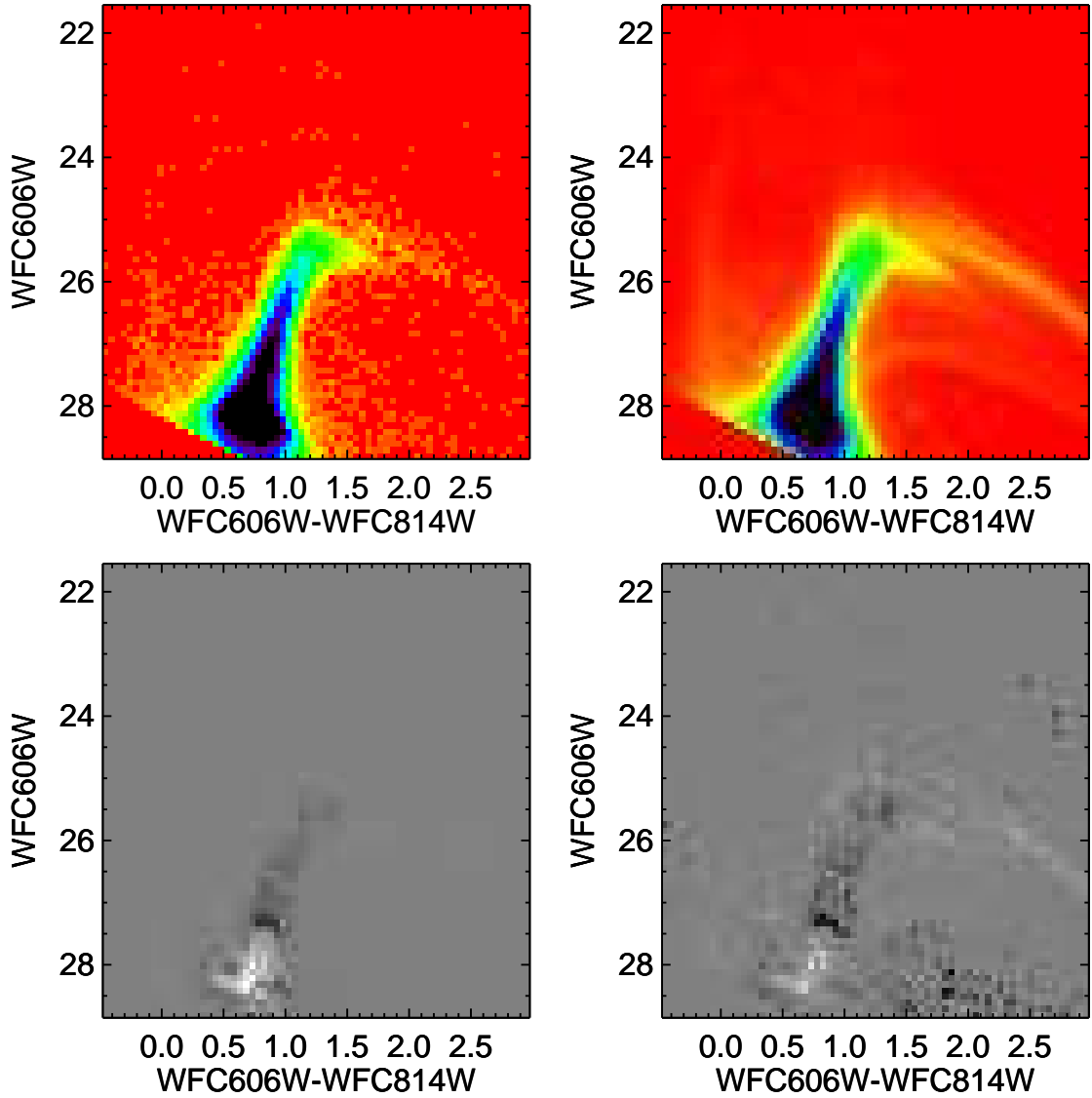,width=6.0in,angle=0}}
\caption{Our best full CMD fit to the data from the OUTER field (see
\S~\ref{fitting}).  {\it Upper-left:} The observed CMD. {\it
Upper-right:} The best-fitting model CMD from MATCH. {\it Lower-left:}
The residual CMD (data-model).  The range plotted is from -163 to +109
stars bin$^{-1}$ (lightest to darkest). {\it Lower-right:} The
deviations shown in {\it lower-left} normalized by the Poisson error
in each CMD bin.  This plot shows the significance of the residuals in
{\it lower-left}.  The most significant residuals correspond to
roughly -21\% and 27\% deviations (lightest to darkest) and occur at
the locations of the red clump (F606W$\sim$28.3; overpopulated by the
models) and AGB bump (F606W$\sim$27.3; underpopulated by the models),
where the current models are known to suffer from deficiencies
\citep{gallart2005a}.}
\label{residuals}
\end{figure}

For all regions, we initially used a fine logarithmic time and
metallicity resolution (0.1 dex) to allow the best possible fit to the
data.  The resulting fit to our deep OUTER field is shown in
Figure~\ref{residuals}.  While the overall fit to the CMD is quite
good (upper panels), there are significant discrepancies in the red
clump (overpopulated by the models) and brighter AGB bump
(underpopulated by the models; see lower panels).  Current isochrone
models are not able to perfectly reproduce these features, which is a
well-known deficiency of the isochrone set \citep{williams2008}.
After performing the full CMD fit, we binned the results to coarser
time resolution to reduce our SFH uncertainties and to avoid drawing
conclusions based on details that could be affected by deficiencies in
the models or could not reliably be constrained by the data \citep[see
discussion in \S4 of ][and \S2.4.2 below]{williams2008}.

\subsubsection{Uncertainty Estimates}\label{uncertainty}

\begin{figure}
\centerline{\psfig{file=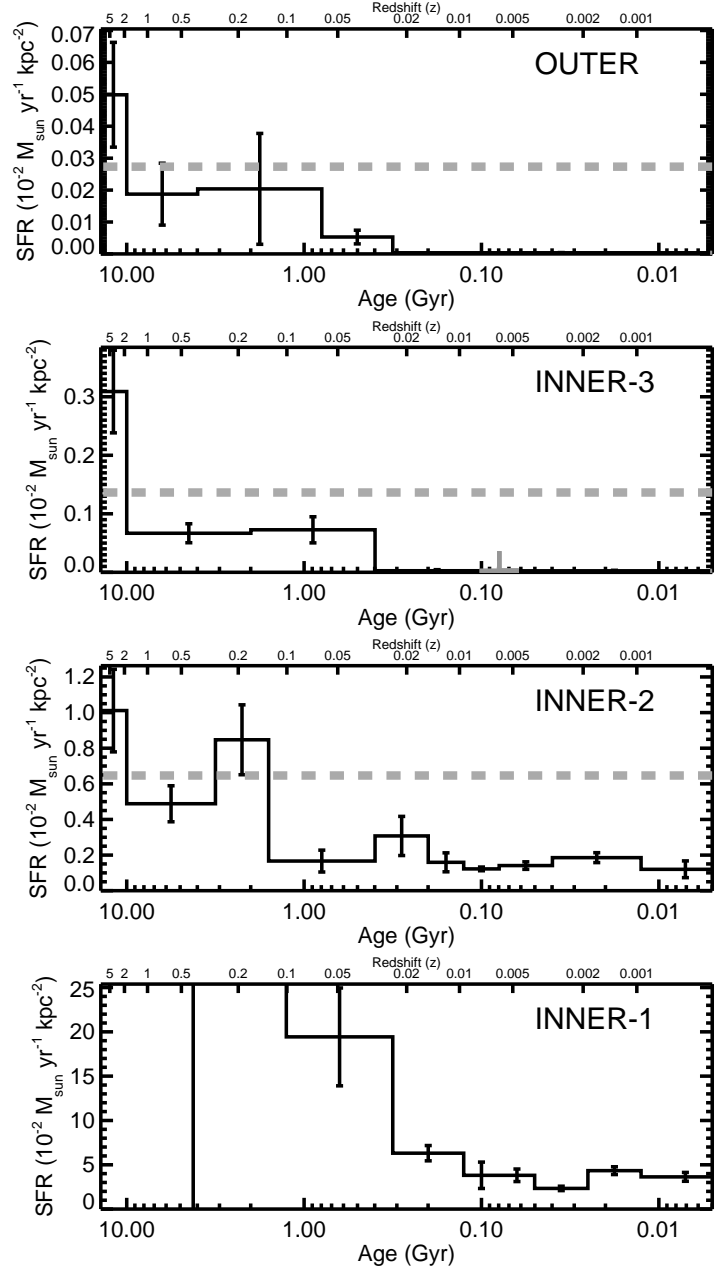,width=3.7in,angle=0}}
\caption{The SFH of the outermost (top) to innermost (bottom) regions
(see \S~\ref{timebins} and \S~\ref{results}). The solid histogram
marks the star formation rate surface density as a function of time
for the past 14 Gyr. The dashed line marks the mean rate.  Edges of
time bins with rates too low to see are marked with heavy gray ticks
on the bottom axis.  The outer 3 regions show star formation that
declines to the present, with star formation truncating more than 300
Myr ago in the outermost field. The innermost field was too shallow to
provide any constraint on the population older than 3 Gyr (error bar
extends off the plot in both directions for the oldest time bin).}
\label{sfr}
\end{figure}

\begin{figure}
\centerline{\psfig{file=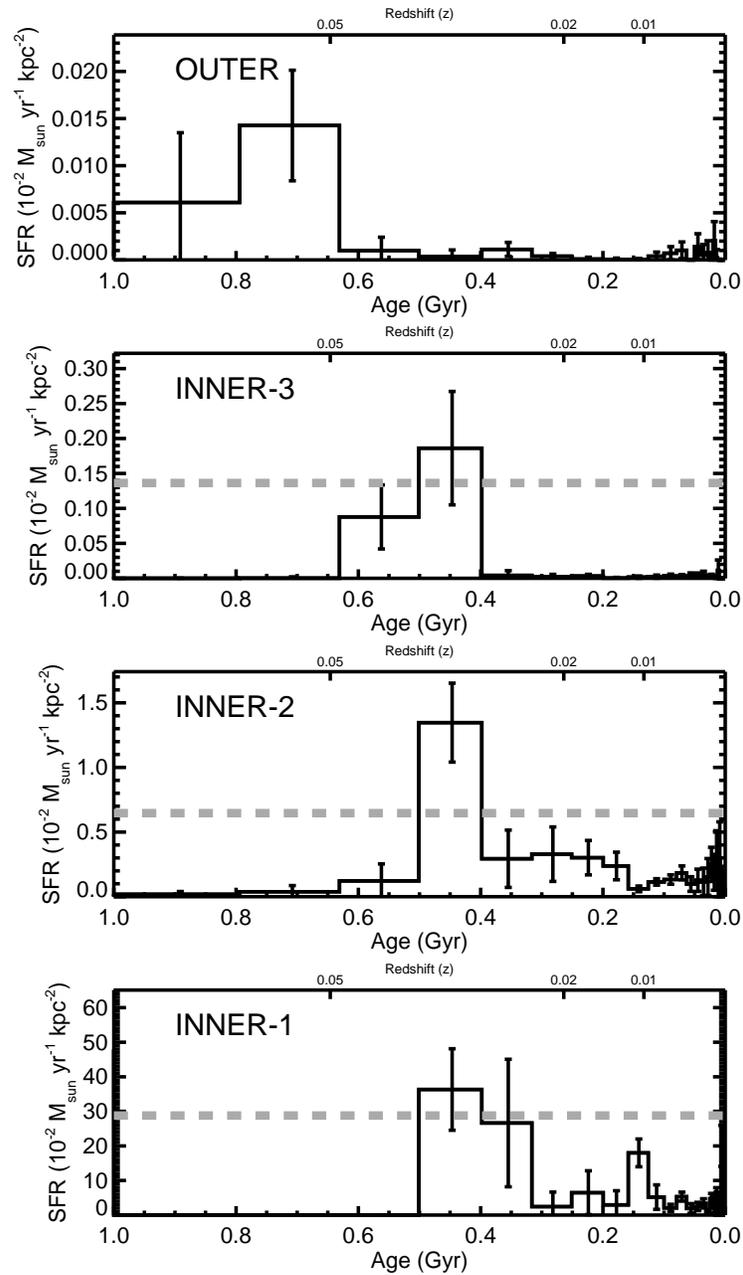,width=3.7in,angle=0}}
\caption{Recent SFHs for the 4 regions for the past Gyr plotted on a
linear timescale at 0.1 dex time resolution (see \S~\ref{discussion}).
Differences between these and Figure~\ref{sfr} reveal the effects of
applying time resolution that is finer than the data can probe, as
determined by our Monte Carlo tests (see \S~\ref{timebins}).  The
dashed line indicates the mean rate for the full history of the
galaxy.}
\label{recent_sfr}
\end{figure}

\begin{figure}
\centerline{\psfig{file=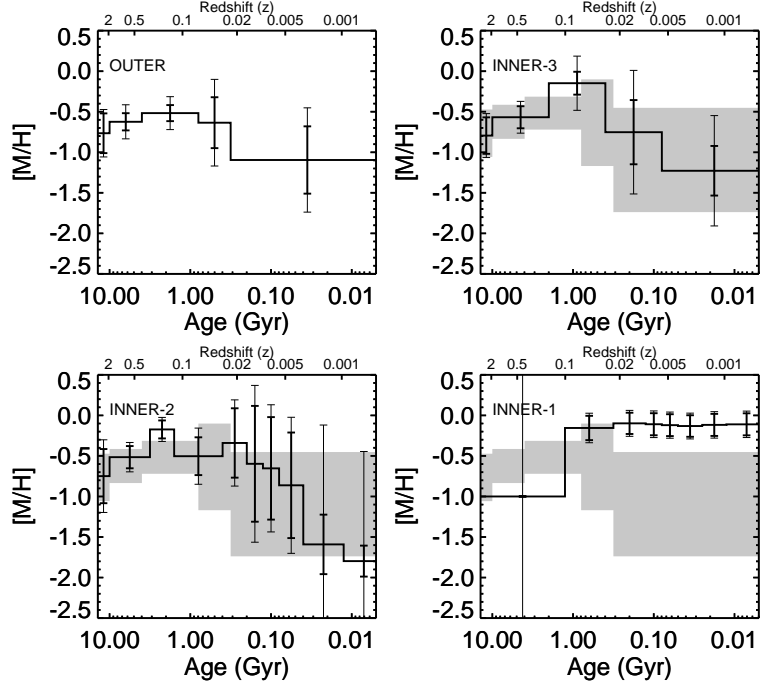,width=4.0in,angle=0}}
\caption{The mean metallicity and metallicity range of the population
as a function of time for each of the regions (see~\S~\ref{outer}).
Heavy bars mark the measured metallicity range, and lighter error bars
mark how that range can slide because of uncertainties in the mean
metallicity.  There is little significant evidence for any evolution
in metallicity, despite on-going star formation.  Shaded regions show
how the metallicity history measurement from our deepest data (OUTER)
compares with the measurements from the shallower inner regions.  Our
data do not put tight constraints on the metallicity at young ages
($\lap$100 Myr).  Errors are smaller and the mean metallicity is
different for the innermost region because of our assumption of
increasing metallicity with time when fitting these shallower data.}
\label{z}
\end{figure}

\begin{figure}
\centerline{\psfig{file=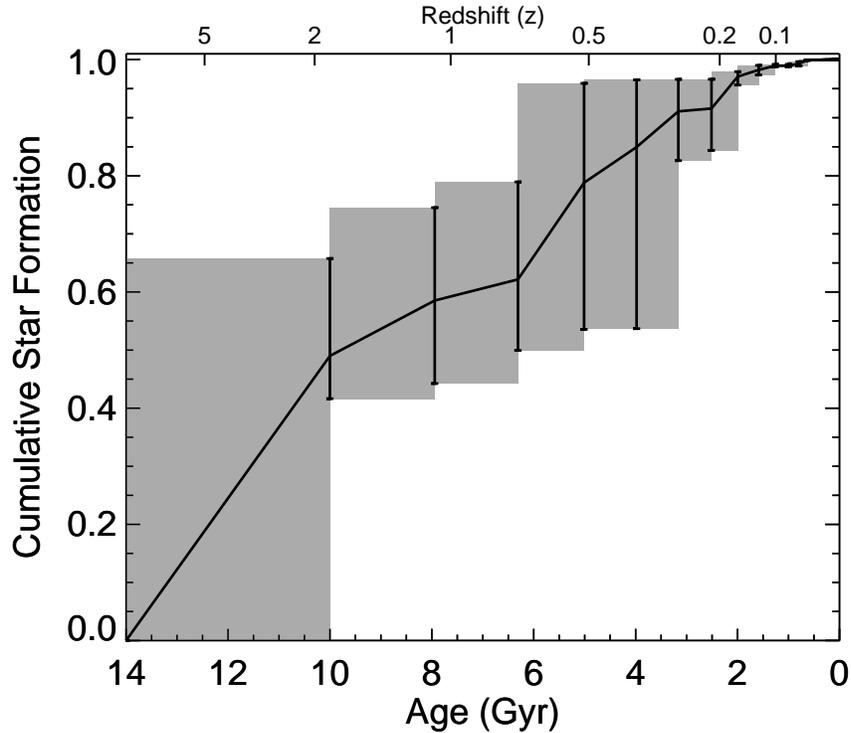,width=5.0in,angle=0}}
\caption{ The normalized cumulative star formation of the OUTER field
as determined by the MATCH package (see~\S~\ref{outer}). {\it Black
Line:} the best fit, assuming constant star formation within each time
bin. {\it Gray Regions:} the evolution allowed by the measured errors.
Roughly 60\% of the stellar mass was formed by $z=1$.}
\label{cum}
\end{figure}

The SFHs shown in Figures~\ref{sfr}-\ref{cum} are subject to a number
of uncertainties.  These include both systematic uncertainties due to
distance, extinction, and deficiencies in the model isochrones, as
well as random errors due to the limited number of stars sampling the
CMD features and the quality of the photometry.  Estimates of the
systematic uncertainties in color and magnitude in the models are
automatically calculated by MATCH, which assesses the fit quality for
a range of distance and extinction values.  This process simulates
uncertainties in the models, which are the systematic uncertainties in
color and magnitude.  We estimate the amplitude of the random
uncertainties by generating 100 CMDs by randomly drawing stars from
our observed CMD, allowing each bin in the Hess diagram to be
populated according to a Poisson distribution, and measuring SFHs for
the resulting CMDs.  We adopted the standard deviation of these SFHs
as our the random SFH errors for each subfield.  These errors were
then added in quadrature to the systematic errors determined by MATCH
from fits using the range of possible distance and reddening values.
The total gives our final uncertainties on the rate, metallicity, and
cumulative fraction of stars formed as a function of time.  We note
that while these uncertainties completely describe the quality of the
data, they do not account for covariance between adjacent time bins
\citep{holtzman2009} and cannot quantify unknown uncertainties,
such as evolutionary phase lifetimes, in the models.

\subsubsection{Assessing Temporal Resolution}\label{timebins}

We have chosen time bin widths to reflect our sensitivity to age. To
measure this sensitivity, we developed an iterative procedure to
adjust the time bins included in the fit. Our initial fits to the CMDs
use fine time bins of 0.1 dex.  We then test the sensitivity to each
of these 0.1 dex bins by removing each one and refitting the data. If
a bin could be removed without significantly reducing the quality of
the fit, we did not include it as a single bin in the final SFH, and
we combined the bin with adjacent bins.  We then iteratively removed
adjacent 0.1 dex bins from the fitting until the quality of the fit
was significantly worse.  We define a significantly worse fit using
the distribution of fit values from our Monte Carlo test runs.  We
calculate the standard deviation of this distribution, and we consider
a fit values more than one standard deviation away from the best fit
to be significantly worse.

However, this technique was biased against time bins that may truly
contain no star formation, since such bins can never have an impact on
the CMD regardless of their duration. Therefore if we found a time bin
(or set of adjacent time bins) whose exclusion from the fit did not
significantly change the fit quality, but that was surrounded by time
bins whose exclusion from the fit did significantly change the fit
quality, we allowed the time bin to remain independent, as a possible
quiescent period in the SFH.  For example, consider 3 adjacent time
bins A, B, and C, and assume the exclusion of time bin B had no
significant impact on the fit.  If the exclusion of bin A negatively
affected the fit and the exclusion of bin C also degraded the fit,
then we infer that B is within a time period that is well-sampled in
the CMD.  In such a case, we do not merge bin B with bin A or bin C,
even though B may not contain any star formation.  In this way, we
allow bins to contain no star formation.

The results of our time bin determination tests were qualitatively
what would be expected.  For the shallowest photometry of the INNER-1
region, we have only two statistically meaningful age bins for all
stars $>$300~Myr.  With the dust and crowding problems in this portion
of the galaxy, the He-burning sequences blend with the RGB at
$M_{F814W}\sim-3$, which corresponds to ages of $\sim$200~Myr.
Furthermore, with the overlapping features and spread in RGB color
from both the large photometry errors from crowding and significant
internal extinction, the detailed age distribution of older stars from
the RGB should not be reliable, as indeed our time bin tests reveal.

For the less crowded and less dusty regions of the INNER field, we
found that we could use 2 age bins for ages $>$2 Gyr (2--10~Gyr and
10--14~Gyr). The independence of these two bins may be due to
variations in the ratio of AGB to RGB stars and the slope of the RGB.
A younger RGB ($<$10 Gyr) is slightly bluer, has a slightly steeper
slope and a higher AGB/RGB ratio than an older RGB ($>$10 Gyr).

For the deep OUTER photometry, there is more information encoded in
the CMD due to the presence of the older stellar populations that
dominate the red clump.  These features provide additional constraints
on the SFR and allow shorter bin divisions at ages $>$1 Gyr.
 
The uncertainties in the derived metallicities tend to be largest at
young ages ($<$100 Myr), where the only metallicity information comes
from the short-lived He-burning sequences.  These features tend to
contain a relatively small number of stars, and the effects of
metallicity on these features is not well understood in stellar
evolution models \citep[e.g.][]{maeder2000,gallart2005a,przybilla2009}.

Overall we are able to obtain very reliable estimates of the relative
contributions of stars of old ($>$10~Gyr), intermediate (2--10~Gyr),
and young ($<$1~Gyr) ages despite unavoidable sources of uncertainty.
We likewise have reliable metallicities covering all but the youngest
ages and high time resolution at young ages ($\lap$300 Myr).  The
cumulative age distribution is particularly stable against the
uncertainties at intermediate ages, because the star formation rates
in adjacent time bins are typically anti-correlated such that some
fraction of the star formation will move back and forth between
adjacent bins depending on small changes in the overall solution.
Therefore we plot the cumulative distribution at the full resolution
of the CMD fit.

\section{Results}\label{results}

The SFHs of all of our regions are shown in
Figures~\ref{sfr}-\ref{cum}.  We first describe the most notable
features of the results for each region, before discussing the
integrated evolution of the galaxy in \S~\ref{discussion}.

\subsection{The SFH of OUTER}\label{outer}

The outer disk of NGC~2976 appears to be dominated by an old ($\gap$8
Gyr), intermediate-metallicity (\hbox{-0.5$\lap$[M/H]$\lap$-1 -0.5})
population, much like other outer disks, thick disks, and inner halos
of ellipticals \citep[e.g., see summary figure in][]{williams2008}.
About 60\% of its outer disk stars were formed by $z\sim1$ (see
Figure~\ref{cum}).  This age is consistent with the mean age of the
M33 outer disk \citep{barker2007}, which has a similar mass
(V$_c\sim110$ km s$^{-1}$ vs. V$_c\sim85$ km s$^{-1}$ for NGC~2976).

While the SFH of the outer disk of NGC~2976 is consistent with roughly
constant star formation over much of the age of the universe, in the
last Gyr there has been a significant decline.  Indeed, its recent
star formation rate has been more than a factor of 5 lower than
average for the past $\sim$800 Myr and more than a factor of 50 lower
than average over the past $\sim$300 Myr.  No areas of current star
formation are apparent in the GALEX UV image \citep{paz2007} or in
deep H$\alpha$ images from SINGS \citep{kennicutt2003}, which is
consistent with our measurements. In fact, the mean age of the stars
formed in the past Gyr is 740$\pm$150 Myr, and no stars younger than
130 Myr are required to produce an acceptable fit to the observed CMD
(as determined by the statistical time bin tests described in
\S~2.4.2).

\begin{figure}
\centerline{\psfig{file=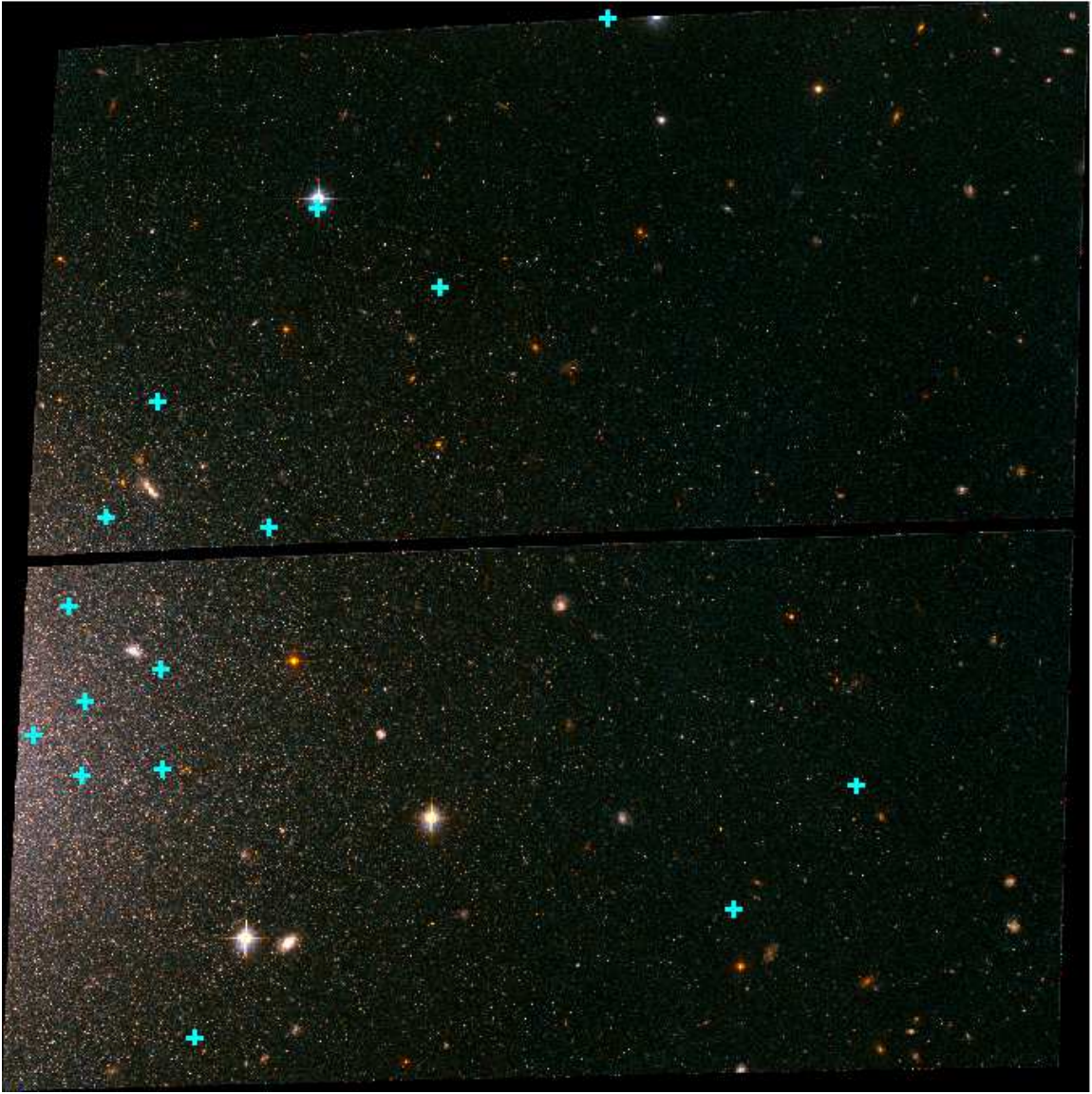,width=6.0in,angle=0}}
\caption{A color image of our DEEP outer field.  Cyan crosses mark the
locations of objects that fall on the upper main sequence in our CMDs.
No clustering of the most massive young stars is detected in this
portion of the outer disk according to a 2-d Kolmogorov-Smirnov test
(see \S~\ref{outer}).  We verified that our quality cuts did not
reject stars in clusters.}
\label{xy}
\end{figure}

Although there are few young stars present in the outer disk, there is
some question as to their origin.  These stars may have formed {\it in
situ}, or they may have scattered from the inner disk.  If they formed
in the outer disk, they may show some signs of clustering. However, as
we show in Figure~\ref{xy}, where we plot the spatial distribution of
upper main-sequence stars in this field (defined as $24<F814W<26$ and
$-0.25<F606W-F814W<0.05$, or equivalently
$5\,M_{\odot}\lap\,M\,\lap12\,M_{\odot}$), these young stars are not
found in clusters outside of the inner disk (seen in the lower-left
edge of Figure~\ref{xy}).  The youngest stars appear smoothly
distributed.  A 2-d Kolmogorov-Smirnov does not show a significant
difference between the distributions of the upper main sequence stars
and the red giants in the OUTER field (only 74\% probability that the
parent distributions are different).

We believe that the most plausible origin for the youngest stars we
observe in these outer regions is that they were born in very small
clusters that dissolved on short timescales. Typical small clusters
with low star formation efficiency dissolve on timescales $\ll$100 Myr
\citep{lada2003}, easily short enough to account for the few young
stars we observe. However, we cannot conclusively determine if their
birth clusters were within the outer disk.  Lower-mass stars formed in
the same clusters may still be present, but cannot be uniquely
identified as young unlike the upper main sequence plotted in
Figure~\ref{xy}.  

As somewhat less plausible alternatives, the lack of current star
formation or clustering of the few massive young stars ($\lap$300~Myr)
in the outer disk is also consistent with the possibility that the
young stars were not formed in the outer disk at all, but instead were
scattered there from supernova kicks \citep[e.g.,][]{hoogerwerf2001}
or interactions with the large number of star forming regions in the
inner disk \citep{roskar2008}.  If these stars migrated the $\sim$2
kpc from the INNER-2 region in 130~Myr, their mean outward velocity
would have been $\sim$14 km s$^{-1}$.  Comparable mean outward
velocities are seen in disk formation simulations, which show stars
rapidly moving from circular orbits in the inner disk to circular
orbits in the outer disk \citep{roskar2008}.

\subsection{The SFH of the INNER Regions}\label{inner}

With the shallower photometry of the inner regions, we cannot reliably
probe the detailed age distribution of the ancient populations,
because the red giant branch alone can be equally well-fit by multiple
combinations of ages and metallicities when there is no constraint on
the fit from fainter CMD features, such as the red clump.  This
problem is often known as age-metallicity degeneracy, which is
difficult to break with such shallow photometry.  However, we still
have ample information on SFRs during the past Gyr, which populates
the luminous main sequence and He burning sequences.  We therefore
will focus most of our discussion of the inner regions on the recent
SFH.

Although in the discussion below we present results for the
F606W-F814W filter combination, we note that we have independently
verified the recent SFHs ($<$500~Myr; $<$300 Myr for the INNER-1
region) using the F475W-F814W CMDs.  In all 3 cases, the SFHs measured
from the F475W-F814W CMDs were consistent with those measured from the
F606W-F814W CMDs in this age range.

\subsubsection{The SFH of INNER-3}\label{inner3}

Like the OUTER region, the SFH of this field shows no evidence for
significant recent star formation.  This consistency is not
surprising, as there is substantial overlap in the projected radii of
stars in these regions (see Table~\ref{depth}).  After 400~Myr ago,
there is a dramatic decrease in the star formation rate.  The mean age
of the stars formed in the past Gyr is 484$\pm$58 Myr, and no stars
younger than 100~Myr are required to produce an acceptable fit to the
observed CMD, as determined by the statistical time bin tests
described in \S~2.4.2.

Together, the star formation shutoff times and the mean ages of the
young populations in OUTER and INNER-3 hint that we may be measuring a
difference in the populations of these regions. However, this
difference turns out to be attributable to the different depths of
photometry in the two regions.  When we remeasure the SFH of the OUTER
region without including the photometry of stars fainter than
F814W=26.8, the same cutoff used when fitting the INNER-3 data, the
measurement of the mean age of the stars formed in the past Gyr drops
somewhat to 580$\pm$130~Myr and the age of the youngest stars required
to fit the CMD drops to 100~Myr.  This test suggests that INNER-3 and
OUTER contain similar populations; however, our INNER-3 photometry
lacks the depth necessary to determine if there is a significant age
difference between the young populations of the 2 regions.

Although we do not have very reliable measurements of the old
populations in this regions, we note that our fits suggest the age
distribution prior to the steep drop in star formation is consistent
with being constant from $\sim$1--10 Gyr, although variations on
shorter timescales are certainly allowed by the data, given the large
width of our time bins in this region. The age and metallicity
distribution of the old stars in this region appears comparable to
that of all the regions studied, including the much deeper OUTER
region.

\subsubsection{The SFH of INNER-2}\label{inner2}

This region shows only a small decline in star formation $\sim$200 Myr
ago.  Unlike the complete truncation of star formation seen in INNER-3
and OUTER, stars have continued to form at this radius, albeit at a
reduced rate.  The mean age of the stars formed in the past Gyr is
somewhat younger than that of the INNER-3 region (394$\pm$42 Myr), and
the youngest stars required to fit the CMD are also younger (13~Myr).
This age difference is consistent with that inferred from visual
inspection of the CMDs of INNER-2, which has a much more pronounced
main sequence than that seen in INNER-3 (see Figure~\ref{cmds}).
Furthermore, it is consistent with the truncation of star formation
from the outside-in.

Furthermore, the old population appears again similar to those of the
other regions.  This similarity of the ancient populations across such
a large radial extent may indicate that the older population is more
well-mixed than the younger population, as would be expected given the
relatively short dynamical time for the galaxy ($\sim$180 Myr) and
current barred kinematics \citep{spekkens2007}. While the continuous
exponential surface brightness profile may favor a pure stellar disk,
one could also interpret the difference between the old and young
populations as being two kinematically distinct components of the
galaxy.  There indeed have been previous claims that NGC~2976 consists
of a young disk and an old spheroidal halo \citep{bronkalla1992}, and
the consistency of our measurements of the older populations across
NGC~2976 likewise supports those claims.  Kinematic tests would be
required to verify this scenario, however.

\subsubsection{The SFH of INNER-1}\label{inner1}

\begin{figure}
\centerline{\psfig{file=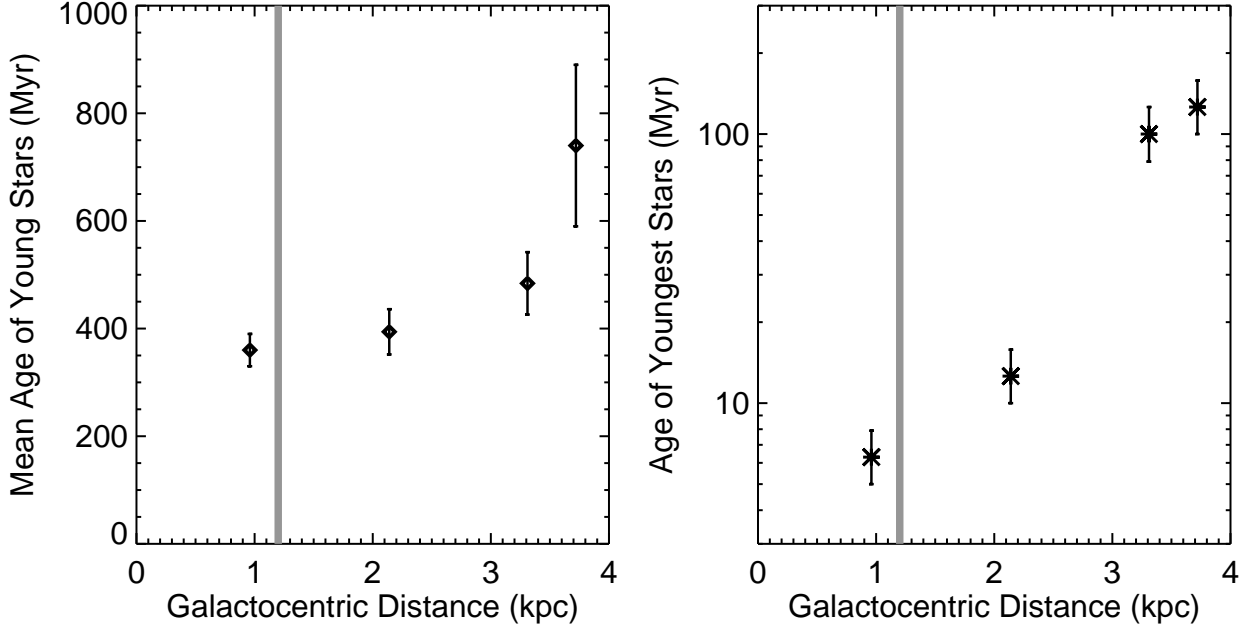,width=6.5in,angle=0}}
\caption{{\it Left:} The mean age of the stellar mass formed in the
past Gyr as a function of median distance from the galaxy center.  The
age of the young stars increases with galactocentric distance. {\it
Right:} The age of the youngest stars necessary to produce an
acceptable fit to the CMD as a function of median distance from the
galaxy center (see \S~\ref{discussion}). Errors mark our 0.1 dex time
bins.  {\it Gray vertical lines:} The location of the disk break. By
both measurements, the age of the young stars increases monotonically
with galactocentric distance.}
\label{age_vs_r}
\end{figure}

The SFH of the innermost region is the most difficult to constrain due
to its high crowding, bright completeness limit, and significant
differential extinction.  These effects cause confusion in the CMD at
magnitudes as bright as F606W$\sim$24.  These limitations reduce the
reliability of the age distribution for ages $\gap$300 Myr ago and
remove any constraint on the SFH prior to 3 Gyr ago. However, the mean
age of the stars formed in the past Gyr in this region is 360$\pm$30
Myr, again younger than those of the other regions.  In addition, the
youngest stars necessary to fit the CMD are 6~Myr.  These results show
that the age of the youngest stars and the mean age of the stars
formed in the past Gyr increase monotonically with galactocentric
distance (see Figure~\ref{age_vs_r}).  Together, all of the data are
consistent with the truncation of star formation from the outside-in.

\section{Discussion: The Recent SFH of NGC~2976}\label{discussion}

NGC~2976 does not look like any other nearby galaxy.  Its bright inner
disk has little organized structure and a sharp truncation edge.  The
atypical appearance of NGC~2976 could indicate that this galaxy is in
the midst of a short-lived and interesting time in its evolution.  The
recent shutdown of star formation seen in the SFHs of the outer
regions of the disk is also consistent with the possibility that
NGC~2976 is undergoing a metamorphosis.  To gain additional insight
into this process, we studied the recent SFHs of our 4 regions in more
detail, including detailed comparisons with the galaxy's kinematics
and gas content.

The recent SFHs of our 4 regions are shown with 0.1 dex time
resolution in Figure~\ref{recent_sfr}.  Although these small time bins
are finer than the sensitivity of the data (shown in
Figure~\ref{sfr}), we include it to ease comparisons with our analysis
of the gas consumption below.  There is clearly a lack of recent star
formation in the outer regions as compared with the inner regions,
which we interpret as a shutdown of star formation from the
outside-in.  This shutdown appears to have begun in the outer disk
$\sim$0.5--1~Gyr ago.  There is also an increase in the age of the
young ($<$1 Gyr) stars with galactocentric distance, shown in
Figure~\ref{age_vs_r}, consistent with this interpretation. We return
to the origin of this shutdown in \S~4.2 below.

\subsection{Timescales}\label{timescaletext}

The evolution of the star formation in NGC~2976 could be related to a
number of different timescales in the galaxy, including the dynamical
timescale, the time since the last interaction with M81, and/or the
gas consumption lifetime in the disk. These timescales are provided in
Table~\ref{timescales}.

We estimate the dynamical time using the H~I rotation curve of
\citet{deblok2008}.  NGC~2976 has a rotation velocity of $\sim$85 km
s$^{-1}$ at 2.5 kpc ($\sim$2 inner disk scale lengths), giving it a
dynamical timescale of $\sim$180~Myr, somewhat shorter than the
apparent star formation truncation timescale.

The time since the last interaction with M81 is not well-constrained,
but can be estimated by making some reasonable assumptions.  Even
though NGC~2976 lies outside the core of the group, the H~I structure
of the group suggests that NGC~2976 has had recent interactions with
the galaxies in the core of the M81 group \citep[M81, M82, and
NGC~3077][]{yun1999,appleton1981}. Likewise, its tidal index is
elevated \citep[$\Theta$=2.7;][]{karachentsev2004}, also consistent
with recent interactions.  Assuming that the relative velocity of
NGC~2976 and M81 is not more than the velocity dispersion of the group
\citep[110 km s$^{-1}$;][]{huchra1982} and that their relative
measured distances in the ANGST survey \citep[3.57\,Mpc for NGC~2976
and 3.58\,Mpc for M81,][]{dalcanton2009} are correct, then any close
passage was at least 1.3~Gyr ago assuming an angular separation of 148
kpc.  Our SFHs do not detect any burst in activity near that age
limit.  However, a passage close to our lower time limit could have
potentially stripped the halo gas and/or induced the radial inflow of
gas from the outer disk, causing subsequent star formation to truncate
from the outside-in (see \S4.3).

The gas consumption timescale of the INNER-1 region is comparable to
the truncation timescale we see in our SFHs.  From the THINGS H~I map
\citep{walter2008}, the current mean gas density (assuming
$\Sigma_{\rm gas}$=1.45~$\times$~$\Sigma_{\rm H I}$) of the NGC~2976
inner disk is $\sim$10~M$_{\odot}$ pc$^{-2}$ out to 90$''$
(r$\sim$1.6~kpc), beyond which it rapidly declines to the noise level
($\Sigma_{\rm HI}\sim$~1.5~M$_{\odot}$ pc$^{-2}$) by 200$''$
(r$\sim$3.5~kpc). Our SFH gives a current star formation rate density
in the innermost region of $\sim$0.036~M$_{\odot}$ yr$^{-1}$
kpc$^{-2}$, yielding a gas consumption lifetime of $\sim$280~Myr in
the inner disk.  If we were to normalize our rates to a
\citet{kroupa2002} IMF, the SFRs would be a factor of $\sim$1.5 lower,
making the gas consumption lifetime closer to $\sim$0.6 Gyr.  In
either case, these timescales are similar to that at which the star
formation is measured to be shutting down.

The gas consumption lifetime of the INNER-2 region ($\sim$2.6~Gyr) is
considerably longer than that of the INNER-1 region. This longer
lifetime reflects the lower star formation rate we have measured for
this region.  While star formation in this region has apparently
slowed down, it has not yet ceased.

Outside of the INNER-2 region, the gas is of very low surface density
($\sim$1.7~M$_{\odot}$ pc$^{-2}$), well below the canonical
\citet{kennicutt1989} star formation threshold.  As may therefore be
expected, we measure only an upper-limit on star formation in the most
recent epoch ($<$8$\times$10$^{-6}$~M$_{\odot}$ yr$^{-1}$ kpc$^{-2}$
in the 4--10 Myr bin), which yields a gas consumption lifetime greater
than a Hubble time.  Even if we consider the higher mean rate back to
80 Myr (2.5$\times$10$^{-5}$~M$_{\odot}$ yr$^{-1}$ kpc$^{-2}$), the
gas consumption lifetime is still greater than a Hubble time,
indicating that significant star formation has effectively ceased in
the outer disk.

\subsection{Gas Content}\label{gascontent}

The consistency between the gas consumption timescale and the star
formation truncation timescale of the inner disk suggests that gas
consumption has played a key role in the recent evolution of NGC~2976.
To further investigate this possibility, we performed a study of the
correlation between gas density and star formation in the disk of
NGC~2976 over the past several hundred Myr.  We assume that all of the
stars that formed during a given time interval must have been in the
form of gas at the beginning of that time interval.  When this gas is
added to the gas seen at the present day, we can infer the gas density
in the recent past, and then compare it to the subsequent star
formation rate.  We can do this at increasing lookback times, giving
us a way of probing the correlation between gas and star formation
rate at a range of epochs, but at a single location within the disk.

We begin with the present day gas density measurements from THINGS
data, assuming that $\Sigma_{gas} = 1.45\Sigma_{\rm HI}$.  We then
assume that all stars formed from a previously-existing gas disk that
was in place at the beginning of each time interval.  Our method
therefore requires that the gas disk had a higher surface density in
the past in order to provide the material for all of the star
formation that subsequently occurred.

We limit our estimates to the times for which we have the most
reliable SFHs in all regions ($<$500 Myr ago). We also make a
correction for stellar evolution, which returns some fraction of the
new stellar mass back into the interstellar medium, such that a given
mass of new stars consumes a somewhat smaller net mass of gas.  We
assume that stellar evolution quickly recycles 20\% of the mass of
stellar mass back to the gas phase.  This fraction corresponds to all
stars $>$8~M$_{\odot}$ (assuming a Kroupa IMF integrated from 0.1
M$_{\odot}$ to 100 M$_{\odot}$), which return their mass to the ISM on
timescales shorter than 100 Myr.  The exact value of this fraction had
minimal impact on our estimates. 

\begin{figure}
\centerline{\psfig{file=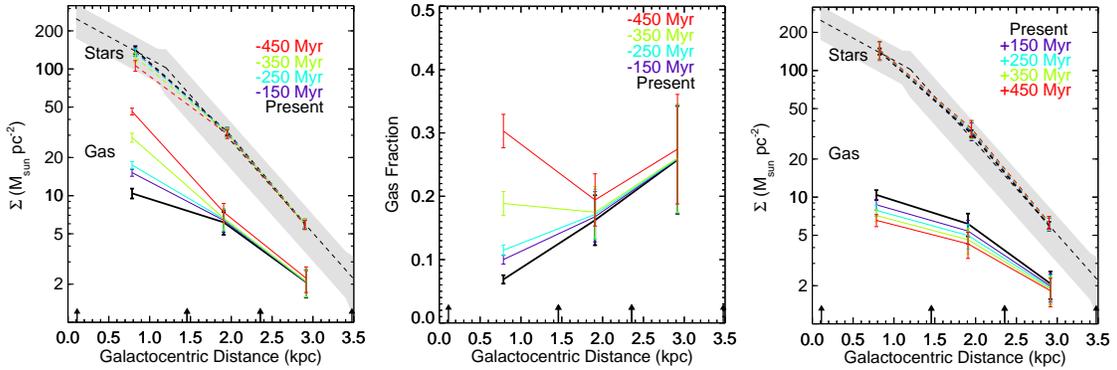,width=6.0in,angle=0}}
\caption{{\bf \it Left:} {\it Solid Lines:} The gas surface density
inferred for different epochs [150 Myr ago (blue), 250 Myr ago (cyan),
350 Myr ago (green), 450 Myr ago (red)], assuming that the stars
formed in each epoch depleted an existing gas reservoir (see
\S~\ref{gascontent}).  The current surface density of gas was measured
from the THINGS data assuming a correction factor of 1.45 to convert
from H~I density to total gas density (black). {\it Dashed Lines:} The
surface density of stars over the past 450 Myr.  Colors mark the same
epochs as for the gas.  Present star density (shaded region) assumes
and extrapolates the \citet{simon2003} K-band profile and $M/L_K$=1.1,
obtained by correcting \citet{bell2003} to a true \citet{salpeter1955}
IMF. Error bars shown are relative errors between the mass densities
at different lookback times; these errors do not include the error in
the original stellar mass profile. {\it Arrows:} Limits of the radial
bins, as defined based on the SFH measurements and the regions with
good H~I data. {\bf \it Middle:} The gas mass fractions in different
radial bins for the same epochs as {\it Left}. {\bf \it Right:} Same
as {\it Left} but the gas densities plotted are anticipated for future
epochs, assuming that the stars formed in each epoch will deplete the
existing gas reservoir [150 Myr (blue), 250 Myr (cyan), 350 Myr
(green), 450 Myr (red)]. Overall, the predicted profiles do not change
as significantly as the reconstructed ones shown in {\it Left}
indicating that the burst of star formation is ending.}
\label{gas}
\end{figure}

Applying our SFH measurements to the current gas surface density at
different galactocentric distances yields a reconstruction of the gas
surface density profile over the past 450 Myr (Figure~\ref{gas}).  The
inferred gas density profile $\sim$450 Myr ago appears to approach
that of a single exponential with a scale length of $\sim$0.7 kpc, or
$\sim$40$''$, similar to the scale length of the outer stellar disk
\citep[34$''$;][]{simon2003}. If our assumptions of little gas infall
or outflow over the past $\sim$450 Myr are correct, then the gas
profile of NGC~2976 had this simpler, single exponential form in the
past.  The gas density of $\sim$50~M$_{\odot}$ pc$^{-2}$ inferred for
the inner disk 450~Myr ago is very high for a galaxy of this mass
\citep[typically $\lap$10~M$_{\odot}$ pc$^{-2}$][]{swaters2002},
suggesting that there may have been radial inflow of gas from the
outer disk $\gap$500~Myr ago.

The very high gas density 450~Myr ago led to a very high star
formation rate, and rapid consumption of the gas.  Indeed, the regions
inside the disk break have converted $\sim$75\% of the gas present
$\sim$500 Myr ago into stars, indicating very rapid gas depletion at
recent times. In contrast, the low surface density regions beyond the
disk break have converted only $\lap$30\% of the gas present $\sim$500
Myr ago into stars.

The rapid decrease in the gas surface density dramatically dropped the
gas mass fraction of the galaxy.  To determine the gas fraction in the
disk as a function of galactocentric distance and time, we started
with the present-day stellar density from the K-band profile of
\citet{simon2003}, assuming $M/L_K$=1.1 (obtained by correcting
\citet{bell2003} to a true \citet{salpeter1955} IMF to agree with our
SFH calculations).  Our estimate of the current total gas mass
fraction ($\sim$0.19 out to 3.5~kpc) agrees within 10\% of those of
\citet{leroy2008} obtained from {\it Spitzer} and THINGS
($\sim$0.17). The estimated gas fraction at previous epochs changes
much more drastically in the inner regions than in the outer regions.
While the gas fraction has stayed relatively constant at $\sim$0.26
for the past $\sim$500 Myr in the outer disk, the gas fraction of the
inner disk has plummeted from $\sim$0.30 to $\sim$0.07.  These values
could be indicative of a massive old stellar population recently
augmented by a long ($\gap$500 Myr) burst of star formation that is
now in the process of ending. Our measurements of the gas fraction at
previous epochs (back to 500~Myr ago) are consistent with the large
range observed by \citet{geha2006} for galaxies of similar baryonic
mass, suggesting that individual galaxies can likely cycle throughout
the observed range in response to gas infall, redistribution, and
consumption.

\begin{figure}
\centerline{\psfig{file=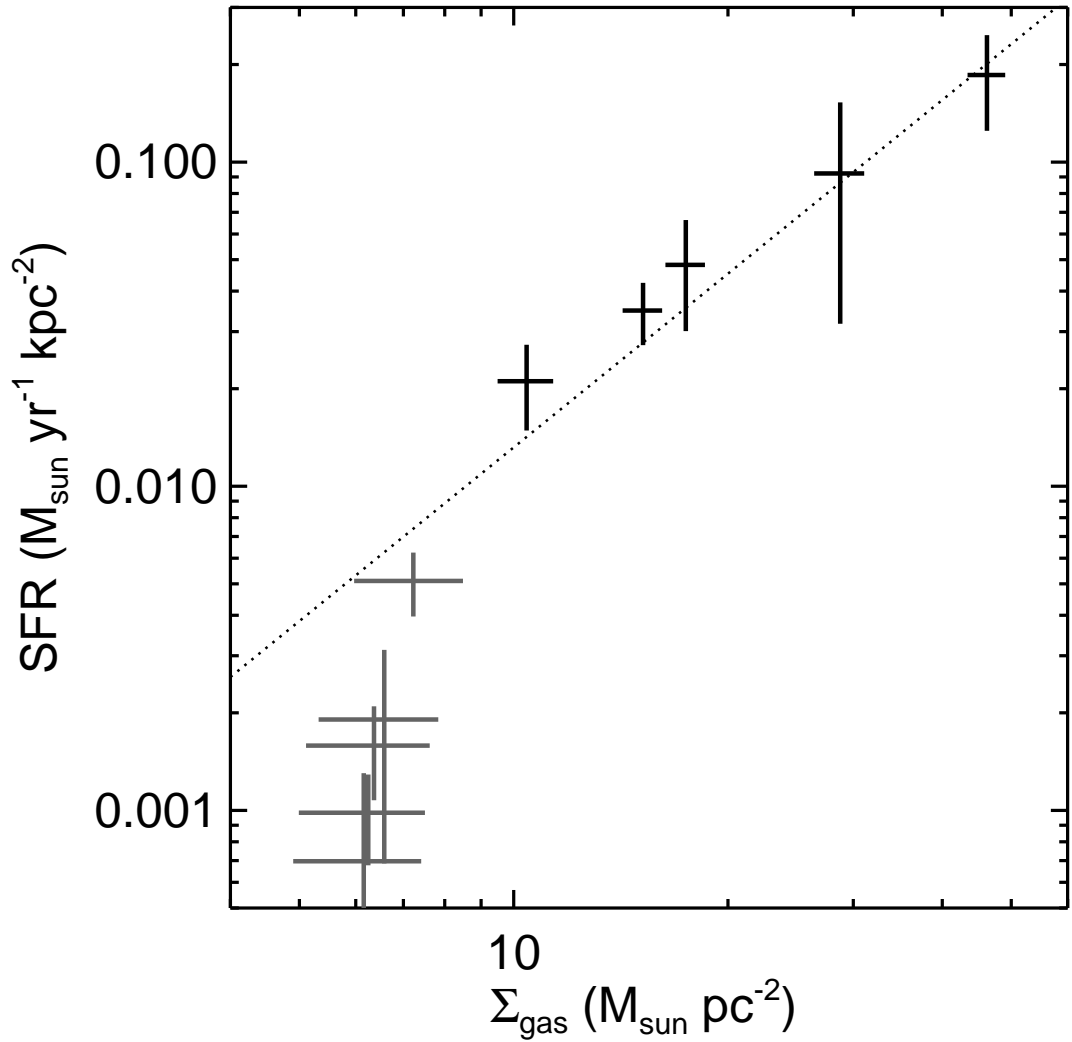,width=6.0in,angle=0}}
\caption{The star formation rate surface density (normalized to a
Kroupa IMF) at previous epochs (back to 450 Myr ago) from our star
formation histories of the INNER-1 (inside the disk break; black) and
INNER-2 (outside the disk break; gray) regions vs. the gas surface
density during those epochs (taken from the left panel of
Figure~\ref{gas}). Within a given region, our assumptions force the
density to increase farther into the past (see \S~\ref{gascontent}).
Therefore higher densities represent earlier epochs, back to 450 Myr
ago with the same time resolution as the left panel of
Figure~\ref{gas}. The dotted line shows the star formation law for
$\Sigma_{\rm HI}$ in NGC~2976 from taken directly from
\citet{bigiel2008}.  Correcting their relation to $\Sigma_{gas}$ would
shift the dotted line to the right.  In general, there is good
agreement with the slope for gas densities $>$7~M$_{\odot}$~pc$^{-2}$.
The reduction in star formation efficiency below
7~M$_{\odot}$~pc$^{-2}$ is comparable to what is seen in other dwarf
galaxies, but steeper than found in THINGS for NGC~2976.}
\label{schmidt}
\end{figure}

Our inferred gas densities at previous epochs also allowed us to study
the historical relationship between average gas density and average
star formation rate density at different radii in the recent past.
Figure~\ref{schmidt} shows the historical relationship between gas
density and star formation rate for the INNER-1 (black) and INNER-2
(gray) regions.  The gas density values are the same as those shown in
the left panel of Figure~\ref{gas}, and the corresponding star
formation rates are those from our SFH measurements averaged over the
same epochs.  Although the axes of this plot are interrelated
(i.e. the cumulative SFH was used to derive the horizontal axis), the
relationship between the differential SFH and the cumulative SFH is
not required to be a power law.

The dotted line in Figure~\ref{schmidt} marks the Schmidt law in the form

$$
\frac{\Sigma_{SFR}}{{\rm M}_{\odot}\  {\rm yr}^{-1}\  {\rm kpc}^{-2}}
=
a(\frac{\Sigma_{gas}}{10\ {\rm M}_{\odot}\  {\rm pc}^{-2}})^{N}.
$$

\noindent
We show the relation found in THINGS for $\Sigma_{H~I}$ in NGC~2976
\citep[log($a$)=-1.88; $N$=1.78,][]{bigiel2008}.  Their gas densities
were determined using the same H~I data used here, but their SFRs were
determined from a combination of GALEX FUV and Spitzer 24$\mu$m maps
and considered only star formation visible at the present day. Their
observed relation was thus based on many regions within the same
galaxy, but only at the present day.

Figure~\ref{schmidt} shows that the historical star formation rates at
the range of gas densities shown in Figure~\ref{gas} are generally
consistent with the present day relationship seen by THINGS.
Therefore the relation measured using a range of SFRs and H~I
densities from different {\it locations} in the disk is consistent
with our values for SFRs and H~I densities from different {\it times}
in the history of the disk.  The consistency suggests that our main
assumption (the gas was in place $\sim$500 Myr ago) is likely to be
correct.  If so, any gas infall and/or inflow responsible for the high
gas densities during that epoch must have occurred prior to 500~Myr
ago, when NGC~2976 may have been significantly closer to the core of
the M81 group.

The observed change in star formation efficiency at low gas densities
(seen as a change of slope in Figure 14) is common in nearby galaxies.
Many of the galaxies in THINGS sample show a steeper relation at gas
densities $\lap 10~M_{\odot}$ pc$^{-2}$ than at higher densities.  The
results from the INNER-2 region show a very steep relation.  This
region, which lies just outside the disk break, therefore has likely
experienced a significant drop in star formation efficiency recently,
as its gas surface density dropped below $\sim$7~M$_{\odot}$
pc$^{-2}$.  This sharp drop in star formation efficiency at low
surface densities is even steeper than that found by THINGS for
NGC~2976.

Finally, we use the relation between gas density and star formation
rate in NGC~2976 measured by \citet{bigiel2008} for gas densities
below 10~M$_{\odot}$ pc$^{-2}$ to predict the future gas and stellar
mass profiles.  Assuming all of the stellar mass that forms depletes
the gas reservoir, we calculate the radial profile of the gas and
stars for future epochs and plot them in Figure~\ref{gas}.  The
stellar mass profile is expected to undergo very little change, and
the gas continues to be depleted while maintaining the shape of its
density profile.  Overall, much less dramatic evolution is seen in the
future prediction than in the past reconstruction, which is again
consistent with NGC~2976 being at the end of a burst of star
formation.

\subsection{Comparisons with Simulations}\label{comparisons}

The young stellar component of NGC~2976 appears to be confined to a
disk.  Although the disk has relatively chaotic structure and a sharp
break, there is only limited evidence that processes and structures
similar to those we observe in NGC~2976 exist in current simulations
of disk formation and evolution.

In simulations of dwarf disks, \citet{stinson2009} have identified a
mechanism by which the gas disk shrinks over time due to the depletion
of gas in the central portions of the disk, which reduces the amount
of available pressure support.  As the gas disk shrinks, star
formation recedes from the outside-in, in qualitative agreement with
our results for NGC~2976.  In their simulations, the process occurs
over longer timescales ($\sim$Gyrs); however, their simulations are
for somewhat smaller disks than NGC~2976, do not include interactions,
and show that the timescale of the process decreases as the mass of
the disk increases.

On the other hand, in simulations of larger disks, \citet{roskar2008}
have shown that radial scattering can lead to a reversal of the disk
age gradient outside of the disk break.  This effect also produces a
gradient in the age of the youngest stars, as it takes longer for
stars to be scattered to larger radii.  This gradient in the age of
the youngest stars is comparable to what we have observed in NGC~2976.
However, the timescale is again longer than that observed in NGC~2976
($>$1\,Gyr predicted vs.\ $<$1\,Gyr observed).

To understand the recent evolution of NGC~2976, comparisons with
simulations and observations of interacting galaxies may be more
appropriate.  It is possible that NGC~2976's recent interaction with
the core of the M81 group caused internal gas redistribution due to
tidal forces and/or an episode of halo gas stripping.  Either of these
processes can lead to a shutdown of star formation from the
outside-in.  

The strong tidal forces within an interaction could cause a major
redistribution of gas within NGC~2976, such that gas from the outer
disk funnels to the inner disk, increasing the central star formation
rate at the expense of the outer disk.  The central metallicities of
interacting galaxies strongly suggest that such radial gas inflow
occurs in such interactions \citep{kewley2006}.  Hydrodynamical
simulations are in general agreement with the observations
\citep[e.g.,][]{iono2004}. The central gas surface densities inferred
for previous epochs for NGC~2976 (\S~\ref{gascontent};
$\sim$50~M$_{\odot}$~pc$^{-2}$) are much higher than those typically
seen in Irr galaxies of similar luminosity to NGC~2976
\citep[$\lap$10~M$_{\odot}$~pc$^{-2}$;][]{swaters2002}.  It is
possible that much of this gas fell into the central regions from the
outer disk prior to 500~Myr ago, when NGC~2976 was closer to the core
of the M81 group.

An interaction with the galaxy group could also result in the
stripping of gas from the NGC~2976 halo.  Numerical simulations of
galaxy groups have recently shown evidence that when halo gas is lost
from disks due to ram pressure stripping, the proceeding truncation of
star formation then occurs from the outside-in.  The timescale for
such truncation due to strangulation is $\lap$0.5~Gyr
\citep{bekki2009} and is shorter for less massive galaxies.  These
simulations are therefore consistent with what we observe in
NGC~2976. Either or both of these interaction scenarios (inflow or
stripping) would be consistent with the recent gas consumption in
NGC~2976 as well as simulations of disk galaxies in group
environments.

\section{Conclusions}\label{conclusions}

We have measured resolved stellar photometry of 2 HST/ACS fields
covering $\sim$6 scale lengths of the NGC~2976 disk.  The resulting
color-magnitude diagrams show a radial trend where the fraction of
main sequence stars decreases with increasing radius, even outside of
the most active central region.  Modeling the CMDs to recover the age
distribution of the stars reveals that, while the outer regions have
undergone very little star formation for the past $\sim$500 Myr, the
inner regions have continuously formed stars to the present.  We see
this lack of young stars begin at a radius of $\sim$3~kpc, well
outside of the disk break.  Inside of 3~kpc, the recent star formation
rate is similar to the rate in past epochs. This trend of increasing
age of young stars with galactocentric distance is similar to an age
trend recently measured in the disk of the Large Magellanic Cloud,
also using CMD analysis \citep{gallart2008}.  The trend suggests a
truncation of star formation from the outside-in.

The radially-dependent differences in the young populations are in
contrast to the apparent uniformity of the old populations across the
galaxy, which appear to be consistent with one another in proportion
and metallicity.  It is possible that the old stars are dominated by a
spheroidal component of the galaxy \citep{bronkalla1992} and only the
young stars are truly sampling the disk.  Alternatively, the old stars
may simply be a well-mixed old disk population.  Given the correct
barred potential of the galaxy \citep{spekkens2007}, radial orbits
should have led to significant radial redistribution of stars.

Detailed study of the gas in NGC~2976 shows clear evidence that it has
recently been rapidly depleted. The current gas densities and gas
consumption lifetimes in NGC~2976 suggest that the dominant mechanism
behind the observed age gradient is the quenching of the gas disk.
The lack of clustering of the few young stars at large radii does
admit the possibility that radial scattering may have played a role in
placing some of the young stars into the outer disk.  However, we
consider the shutting down of star formation in the disk from the
outside-in to be the dominant process in the current morphological
transition occurring in NGC~2976.  This process may be related to ram
pressure stripping of the halo gas and/or internal gas redistribution
induced by tidal forces due to the most recent interaction with the
core of the M81 group.

\bigskip

We thank Gregory Stinson for helping to compare the dwarf simulations
to our measurements.  Support for this work was provided by NASA
through grant GO-10915 from the Space Telescope Science Institute,
which is operated by the Association of Universities for Research in
Astronomy, Incorporated, under NASA contract NAS5-26555.

\clearpage

\begin{deluxetable}{lccccccc}
\tablecaption{Properties of the Designated Regions}
\tablehead{
\colhead{{\footnotesize Region}} &
\colhead{{\footnotesize R$_{in}$ (kpc)\tablenotemark{a}}} &
\colhead{{\footnotesize R$_{out}$ (kpc)}} &
\colhead{{\footnotesize R$_{med}$ (kpc)\tablenotemark{b}}} &
\colhead{{\tiny Stellar Mass Fraction\tablenotemark{c}}} &
\colhead{{\footnotesize $F475W_{50}$\tablenotemark{d}}} &
\colhead{{\footnotesize $F606W_{50}$\tablenotemark{e}}} &
\colhead{{\footnotesize $F814W_{50}$\tablenotemark{f}}} 
}
\startdata
{\footnotesize INNER-1} & 0 & 1.5 & 1.0 & 0.86 & 25.9 & 25.8 & 25.0\\
{\footnotesize INNER-2} & 1.5 & 3.0 & 2.1 & 0.13 & 27.4 & 27.1 & 26.6\\
{\footnotesize INNER-3} & 2.4 & 6.2 & 3.3 & 0.03 & 27.9 & 27.9 & 27.2\\
{\footnotesize OUTER} & 2.4 & 8.2 & 3.7 & 0.03 & 27.7 & 28.9 & 28.2
\enddata
\tablenotetext{a}{Deprojected radii.  These overlap because the
ellipticity of the isophotes used to define the regions were not
precisely those expected for standard value for the inclination
(64.5$^{\circ}$) we assumed for the disk.}
\tablenotetext{b}{The median galactocentric distance of the stars in the the region.}
\tablenotetext{c}{The fraction of stellar mass contained between
R$_{in}$ and R$_{out}$ assuming the profile shown in Figure~\ref{gas}.}
\tablenotetext{d}{The 50\% completeness limit of the F475W data.}
\tablenotetext{e}{The 50\% completeness limit of the F606W data.}
\tablenotetext{f}{The 50\% completeness limit of the F814W data.}
\label{depth}
\end{deluxetable}

\begin{deluxetable}{lrrr}\tablewidth{17cm}
\tablecaption{Relevant timescales for NGC~2976}
\tablehead{
\colhead{Galaxy Process} &
\colhead{Timescale in NGC~2976 (Gyr)}
}
\startdata
Dynamical (rotation) & 0.18\\
Recent shutdown of star formation & $\sim$0.5\phantom{0}\\
Gas Consumption of INNER-1 & \ \ $\sim$0.35\\
Interaction with M81 & $\gap$1.3\phantom{0}\\
Gas Consumption of INNER-2 & $\sim$2.6\phantom{0}\\
Gas Consumption of INNER-3 & $\gap$10\phantom{.00}\\
\enddata
\label{timescales}
\end{deluxetable}

\end{document}